\documentclass{aa}  

\usepackage{graphicx}
\usepackage{subfig}
\usepackage{xcolor}
\usepackage{placeins}

\usepackage{txfonts}
\begin{document}

   \title{A new way to measure the distance to NGC1052-DF2\thanks{Based on observations made with ESO Telescopes at the La Silla Paranal Observatory under programme 110.23P4.001.}}


\author{M. A. Beasley\inst{1}\fnmsep\inst{2}\fnmsep\inst{3}
        \and
          K. Fahrion\inst{4}\fnmsep\inst{5}
          \and
          S. Guerra Arencibia\inst{1}\fnmsep\inst{2}
          \and
          A. Gvozdenko\inst{6}
            \and
          M. Montes\inst{1}\fnmsep\inst{2}\fnmsep\inst{7}
          } 

   \institute{Instituto de Astrofísica de Canarias, Calle Vía Láctea, E-38206 La Laguna, Spain\\
              \email{mikey.beasley@gmail.com}
        \and    
            Departamento de Astrofísica, Universidad de La Laguna, E-38206 La Laguna, Spain
        \and 
            Centre for Astrophysics and Supercomputing, Swinburne University, John Street, Hawthorn VIC 3122, Australia
        \and
            Department of Astrophysics, University of Vienna, T\"{u}rkenschanzstra{\ss}e 17, 1180 Wien, Austria
        \and
            European Space Agency, European Space Research and Technology Centre, Keplerlaan 1, 2201 AZ Noordwijk, the Netherlands.
        \and
            Department of Physics, Centre for Extragalactic Astronomy, Durham University, South Road,  Durham DH1 3LE, UK
         \and
            Institute of Space Sciences (ICE, CSIC), Campus UAB, Carrer de Can Magrans, s/n, 08193 08193, Cerdanyola del Valles, Spain
             }

   \date{\today}

 
  \abstract
   {We have employed a novel way to measure the distance to NGC1052-DF2 using the internal stellar velocity 
   dispersions ($\sigma$) of its globular clusters (GCs). 
   We obtained deep (15.1h), R=18,200, Calcium Triplet integrated-light spectra for 10 GCs in NGC1052-DF2 using FLAMES GIRAFFE on VLT.  For five GCs we measure $\sigma$, along with precision velocities for the whole sample. We also present a new photometric analysis based on 40 orbits of archival {\it Hubble} Space Telescope imaging for 16 spectroscopically confirmed GCs.
   Assuming that the NGC1052-DF2 GCs obey the $M_V$ -- log($\sigma$) relation followed by the Milky Way and M31 GCs, the NGC1052-DF2 GCs indicate a distance, $d=16.2\pm1.3$ (stat.) $\pm1.7$ (sys.) Mpc.  By contrast, adopting a literature distance of $d=21.7$ Mpc from forward modelling of the TRGB, the GCs lie above the Milky Way + M31 relation by $\sim0.6$ magnitudes. For a shorter literature distance of 13 Mpc, the GCs fall below the relation by $\sim0.4$ mag.
   At $d = 16.2$~Mpc, we obtain mean dynamical $M/L_V = 1.61\pm0.44~M_\odot/L_\odot$, and median half-light radii, $r_h =3.0\pm0.5$ pc. This is entirely consistent with Milky Way GCs,  with mean $M/L_V = 1.77\pm0.10~M_\odot/L_\odot$, median $r_h =3.2\pm0.6$ pc. For the further distance of 21.7~Mpc, we obtain systematically lower $M/L_V$ ratios ($M/L_V = 1.19\pm0.33~M_\odot/L_\odot$) which could suggest ages of $\sim6$ Gyr assuming canonical mass functions. However, such young ages are inconsistent with a MUSE stellar population analysis of the NGC1052-DF2 GCs which indicates they are $\sim10$ Gyr old.
   For $d = 16.2$~Mpc, coupled with our new photometry, we find that the properties of the GCs in NGC1052-DF2 appear entirely consistent with those in the Milky Way and other Local Group galaxies. In order to reconcile the further distance with our results, a mass function more dwarf-depleted than the Milky Way GCs must be invoked for the GCs of NGC1052-DF2.}


   \maketitle
%

\section{Introduction}

Galaxies are generally thought to form within dark matter (DM) haloes. Although we do not know what DM is actually composed of, there is extensive observational evidence for a gravitational effect that can be explained by DM. Examples include the gas rotation curves in galaxies (e.g., \citealt{Rubin1978}; \citealt{deBlok2008}), galaxy stellar velocity dispersions (\citealt{Faber1976}; \citealt{Cappellari2006}), hot X-ray emitting gas around galaxies and in galaxy clusters (e.g., \citealt{Forman1985}), gravitational lensing of galaxies and galaxy clusters (\citealt{Mandelbaum2006}; \citealt{Leauthaud2012}) and the kinematics of globular cluster (GC) systems around both early- and late-type galaxies (e.g., \citealt{Cote2003}; \citealt{Strader2011}; \citealt{Posti2019}).
All these approaches generally indicate the existence of a gravitational potential in excess of that accounted for by the presence of baryons alone. 

Out to the virial radius, massive galaxies like the Milky Way are thought to have DM-to-stellar mass ratios ($M_{\rm DM}/M*$) $\sim20-100$ (e.g., \citealt{Moster2010}; \citealt{Behroozi2019}). The case is even more extreme for dwarf galaxies, the lowest mass of which may have $M_{\rm DM}/M*\sim10,000$ (\citealt{Simon2019}). From a theoretical perspective, galaxies are generally required to have DM in order for them to collapse; without DM sufficient structure might be hard to form at high redshift \citep{Davis1985}. In addition, fits to the CMB power spectrum from Planck give a remarkably precise estimate of the dark matter density ($\Omega_c h^{²} = 0.120\pm0.001$)\citep{Planck2020}.

This understanding of DM in galaxies has recently been challenged. NGC1052-DF2, also known as KKSG04, PGC3097693, and [KKS2000]04 \citep{Karachentsev2000} is a low surface brightness galaxy thought to lie some 22 Mpc distant in the direction of the massive early-type galaxy NGC 1052. Studies of the kinematics of NGC1052-DF2's GC system \citep{vanDokkum2018_DM} and it's stellar body (\citealt{Emsellem2019}; \citealt{Danieli2019}) suggest that it may have a very low DM fraction, compatible with no DM at all.

In addition to little DM, DF2 appears to have a rather unusual system of GCs. To first order, the "turn-over" (peak) of the GC luminosity function (GCLF) in galaxies has approximately constant absolute magnitude and can therefore be used as a standard candle \citep{Rejkuba2012}. In detail, the peak of the GCLF does in fact vary with galaxy mass, of order of a few tenths of a magnitude, in the sense that dwarf galaxies tend to have have fainter GCLFs \citep{Jordan2007}. However, the GCLF in DF2 has been claimed \citep{vanDokkum2018_GCs} to be more than a magnitude brighter than the GCLF in the Milky Way and indeed, in most other galaxies with well characterised GC systems
(with some possible exceptions, see e.g., \citealt{Romanowsky2024}; \citealt{Li2024}).
Indeed, if DF2 is truly a DM deficient galaxy, with a truly unusual GC system, then it may challenge some of our ideas of how galaxies and their GCs can form. 

An important part of the the puzzle lies in the distance to DF2. \cite{vanDokkum2018_DM} used I-band surface brightness fluctuations (SBF) to measure the distance to DF2 and obtained $19.0\pm1.7$ Mpc (\citealt{Blakeslee2018} also performed an independent SBF analysis and obtained $d=20.4\pm2.0$ Mpc). 
Subsequently, \cite{Trujillo2019} argued that DF2 is significantly closer than 19 Mpc by using a number of methods including the tip of the red giant branch (TRGB) from {\it Hubble} Space Telescope (HST) imaging, SBF and the GC luminosity function. These authors concluded that a distance to DF2 of 13 Mpc was more consistent with the data available. A shorter distance such as this would reduce the tension in $M_{\rm DM}/M*$ since while the dynamical mass of DF2 drops by nearly a factor of 2, the stellar mass decreases by a nearly factor of 4. In addition, the shorter distance brings the DF2 GCLF in line with other galaxies.  

However, in response to \cite{Trujillo2019}, \cite{vanDokkum2018_distance} used a HST-derived TRGB and SBF tied to the NGC 4258 megamaser and found $d=18.7\pm1.7$ Mpc. Later, 
\cite{Shen2021_TRGB_distance} analysed deeper HST data and obtained a new TRGB distance that placed DF2 at an even further distance of $22.1\pm1.2$ Mpc. This has been subsequently revised downwards to 21.7 Mpc (no uncertainty given) by \cite{Shen2023} due to errors in the initial HST image alignment. 

Over the past 6 years, more than 50 papers have discussed some of the possible formation mechanisms of 1052-DF2 specifically (e.g., \citealt{Fensch2019}; \citealt{Trujillo-Gomez2021}; \citealt{vanDokkum2022_bullet} to name a few), while even more works have explored the formation of DM-free galaxies generally.  Clearly, the true nature of this galaxy needs to be understood, independent means to measure its distance are desirable and further examination of its GCs merited.

 Here we employ a novel way to obtain an independent distance to DF2. Individual GCs in the Milky Way, M31 and Local Group dwarfs follow a common absolute magnitude -- velocity dispersion ($\sigma_{\rm gl}$) relation (where $\sigma_{\rm gl}$ refers to the global stellar velocity dispersion, i.e. the velocity dispersion averaged over the entire GC). This stems from the virial theorem; more luminous GCs are more massive (for a given mass-to-light ratio), and are therefore dynamically hotter. Assuming that the DF2 GCs obey the same relation as GCs in the Local Group, then measurement of $\sigma_{\rm gl}$ for DF2 GCs will return a value for $M_V$ independent of the GC distance from the observer. \cite{Beasley2024} (hereafter, BFG24) explored these "GC velocity dispersion distances" (GCVD) and found distance precisions of  $3-5\%$ are potentially achievable for populous GC systems, and distances better than $\sim10\%$ can be obtained for galaxies with few or even only one GC velocity dispersion measurement available. Here, using deep VLT high-resolution spectroscopy, we apply GCVD to the GCs in DF2.

 In a companion paper (Fahrion et al. submitted; hereafter Paper II) we explore the detailed stellar populations and kinematics of the NGC1052-DF2 GCs via a re-analysis of archival MUSE data for this galaxy.
 
\section{Data}
\subsection{Data acquisition and reduction}

Observations of the DF2 GCs were taken with the FLAMES multi-object spectrograph feeding into GIRAFFE in MEDUSA mode (programme 110.23P4.001). We used the HR21 grating covering a wavelength range  $\lambda 8484-9001\AA$ which includes the calcium triplet region, and yields a spectral resolution of $R=18,200$.

A total of $40\times1360$s (15.1 h) integrations were obtained for 12 GCs, comprising 10 previously identified GCs \citep{vanDokkum2018_DM} and two candidates "GCnew1" and "GCnew8" from \cite{Trujillo2019}. 
Target coordinates were taken from \cite{Trujillo2019}. We also allocated 25 fibres to sky regions for sky subtraction. Typical seeing conditions were better than 0.8 arcsec and these observations were taken in grey time. The S/N of the spectra for the two candidates ("GCnew1" and "GCnew8") turned out to be too low (S/N$\lesssim$5.0 pix$^{-1}$) for secure identification and are not discussed further.

These data were reduced using the ESO Reflex pipeline \citep{Freudling2013} and custom python scripts. A standard reduction procedure was followed (wavelength calibration, correction for fibre-to-fibre response, scattered light correction, flat-fielding, cosmic ray removal) and the spectra were extracted using optimal extraction. 
However, at this stage sky was not subtracted. After some experimentation, we found that superior sky subtraction results could be obtained by using \textsc{pPXF} \citep{Cappellari2004} and supplying master skies, as opposed to using the Reflex pipeline sky subtraction. \textsc{pPXF} offers the significant advantage of optimal template matching with the inclusion of scaling of the sky templates. The master skies were constructed from the co-addition of the 25 sky fibres for each of the 40 individual integrations. We also tried selecting N skies closest to each science fibre but this offered no obvious improvement in the sky subtraction. The master skies were then fed to \textsc{pPXF} and the optimal sky template was obtained and subtracted from the individual science spectra. In general (but not always), the best matching sky templates identified by \textsc{pPXF} were those coincident in time with the science spectrum in question.

The individual reduced sky-subtracted 1-d spectra were then corrected to the heliocentric frame and median combined with 3$\sigma$ clipping of outlying pixel values, with the individual spectra weighted by their median fluxes in the continuum regions blueward of 8750$\AA$.

\subsection{Line spread function}

We determined the line spread function (LSF) of the spectra using high S/N skylines. We fit Gaussian functions to $\sim50$ unblended skylines and determined the full-width half-maxima (FWHM) of the lines as a function of wavelength for all the used GIRAFFE fibres. Lines which showed clear asymmetries or other irregularities were rejected.

As a result of this exercise we found a mean full-width half-maximum (FWHM) of the lines of $0.53\pm0.05\AA$ across all fibres considered, and a negligible dependence on wavelength.  We note that the FWHM we measure is slightly higher than that indicated in the headers of the pipeline products measured from arclines (mean FWHM$=0.48\pm0.04\AA$). Since the sky spectra pass through the full optical system of the telescope and instrument, we regard this as a more reliable measure of the true FWHM for our instrumental set-up.

We note that this LSF is required in order to bring stellar model templates to the correct broadening,  as described shortly. In the following we also use Gaia-ESO stars as templates taken using our spectroscopic set-up which we confirmed have the same LSF as our observations.

\subsection{Measurement of radial velocities and velocity dispersions}
\label{Measurementofradialvelocitiesandvelocitydispersions}

\begin{figure}
  \centering
  \subfloat{\includegraphics[width=\hsize]{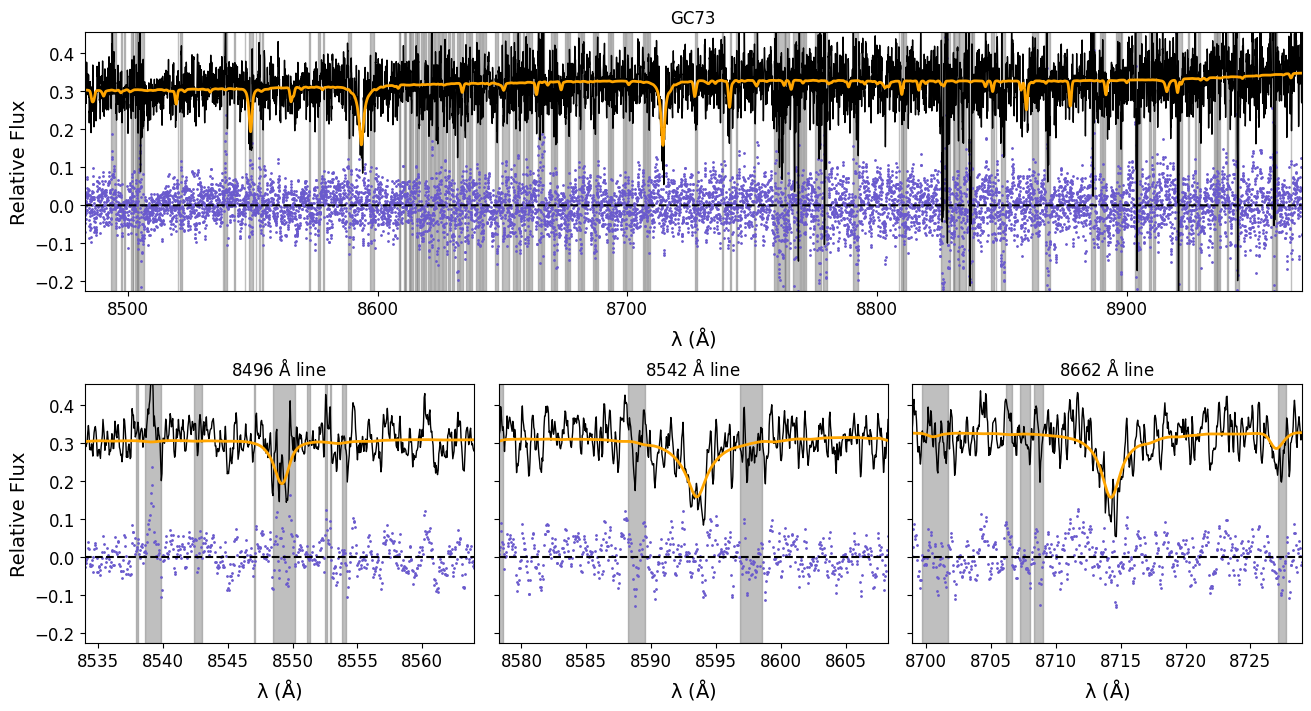}} \\
  \subfloat{\includegraphics[width=\hsize]{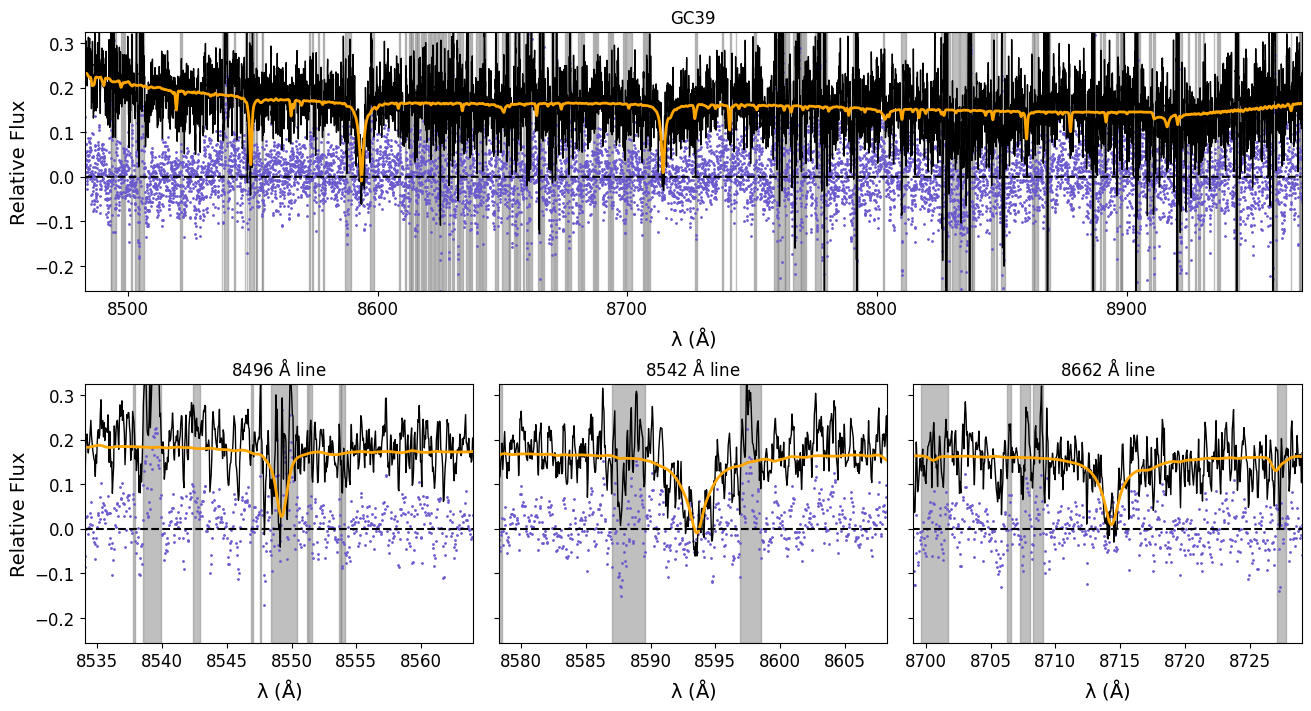}}
  \caption{Example \textsc{pPXF} fits for two GCs for which we measure velocity dispersions (top GC73, bottom GC39). The FLAMES GIRAFFE spectra are shown in black, the optimal model template in orange and residuals of the fit are shown underneath in blue. The greyed out areas are regions of strong sky residuals which are masked out in the fitting procedure. The zoom-in regions are centred around the three lines of the Calcium Triplet. The first line of the Calcium Triplet (at $\sim8549\AA$) is strongly affected by a sky residual. Fits to all the spectra are shown in Fig. \ref{appendix_Fig1} of Appendix~\ref{appendix_A}.} 
  \label{Spectra}
\end{figure}

We fitted the FLAMES GC spectra with \textsc{pPXF} \citep{Cappellari2004}, a full spectrum fitting method widely employed for extracting line-of-sight velocities and dispersions from spectra.\footnote{\cite{Dumont2022} give a good example of the utility of \textsc{pPXF} in deriving velocity dispersions of GCs in Centaurus A.}
\textsc{pPXF} fits the spectrum with a library of suitable model or empirical spectra and we tested the choice of models extensively as described below. 
We chose additive polynomials of degree 4 and multiplicative polynomials of degree 1 to account for variations in the continuum and flux calibrations. 
Due to remaining sky residual lines in the reduced data, we took a careful approach in masking those lines. To do so, we normalised the spectrum without sky subtraction in a continuum wavelength range to the continuum level of the sky subtracted spectrum. Then, we masked out all wavelength pixels in which the normalised sky spectrum exceeds 1.2.
For the high S/N GCs (S/N$\geq$10 pix$^{-1}$) GC39, GC73, GC77, GC85, and GC92, we selected a broad wavelength range between 8530 and 8900 $\AA$, while we restricted the range to 8530 - 8750 $\AA$ for the other GCs due to the strong sky residuals at longer wavelengths. Example fits for GC39 and GC73 and are shown in Fig~\ref{Spectra}, and all our fits are shown in Fig. \ref{appendix_Fig1} of Appendix~\ref{appendix_A}.

To estimate statistical uncertainties, we took a Monte-Carlo approach: after the initial fit, we perturbed the best fitting spectrum with random noise drawn from the residual of original spectrum and best fitting solution and fitted again. This process is repeated 100 times to obtain distributions of the radial velocity and velocity dispersion, from which we extract the 16th, 50th, and 84th percentiles as the best fitting value with its upper and lower uncertainty.

Initially, we used high resolution stellar models from \cite{Coelho2014}. We downloaded\footnote{http://specmodels.iag.usp.br/} the models at a spectral resolution of $R = 20,000$ and sampling of 0.02 $\AA$ per pixel.
For accurate determination of the dispersions, we convolved the models with Gaussian kernels with the LSF as determined previously. From the full library, we selected stars with $4000< T_{\rm eff}<5,000$, $0.5<\text{log}(g) < 3$, and metallicities $< -1$ dex since the spectrum of old GCs in the near-IR is dominated by light from red giant branch (RGB) stars. 

Using these models, we noted that the dispersion could only be measured for the five brightest GCs in the sample (GC39, GC73, GC77, GC85, GC92), whereas measurements of the fainter GCs gave results consistent with zero, showing that the combined limit of the spectral resolution and S/N is reached (see Fig. \ref{appendix_Fig2}). As a consequence, we focused our subsequent analysis on the brightest GCs. 

To establish the robustness of our fits, we tested various different approaches such as varying wavelength ranges, and the degrees of additive and multiplicative polynomials, finding consistent results. However, we noted that the library of models does not reach below a metallicity of $-1.3$ dex, whereas our re-analysis of the MUSE data (Paper II) revealed that the most metal-poor GCs have lower metallicity (e.g., GC77, [M/H] = $-1.61\pm0.09$).

For this reason, we also fitted the GCs with empirical spectra from the Gaia-ESO survey \citep{Gilmore2012}. Here, we selected spectra of Milky Way red giant branch stars with S/N > 100 observed in the same HR21 grating. We selected these stars again based on effective temperature and surface gravity, but also only selected stars within $\pm$ 0.1 dex of the MUSE metallicity of each GC, therefore creating a model library for each of the bright GCs. GC39 is outside of the MUSE field of view and hence has no metallicity measurement, but given its similar colour to that of GC73, we selected stars with $-1.5 \pm 0.2$ dex. While we did not broaden the empirical spectra, we brought them to the same line-of-sight velocity of 1800 km s$^{-1}$. As a consequence, we only used the Gaia-ESO stars to determine the velocity dispersion of the brightest GCs. 

 For additional testing, we used some of the high S/N Gaia-ESO stars mentioned above and fitted them with the same \textsc{pPXF} setup as for the GCs. Firstly, we verified that our fitted velocities match those from the Gaia-ESO catalogue, finding close agreement within 1 km s$^{-1}$. Then, we created mock GC-like spectra out of the Gaia-ESO stars by broadening them with a Gaussian kernel to different values of $\sigma$ and adding noise. We tested how well the initial velocity and input $\sigma$ are recovered as a function of S/N, wavelength ranges, and polynomial degrees. In general, we find that the degrees of the additive and multiplicative polynomials do not affect the results significantly unless extreme values ($>$ 10) are chosen. 
 Testing different input dispersions, we found that they are recovered within 2.0 km s$^{-1}$ for S/N $>$ 10 when the velocity dispersions are $\geq5$ km s$^{-1}$. Dispersions below $\sim$ 4 km s$^{-1}$ are not reliably recovered, likely because of a combination of the intrinsic spectral resolution of HR21 in FLAMES and the limited S/N of the spectra. For S/N$<$10 pix$^{-1}$, we could not reliably recover velocity dispersions irrespective of the value of the input dispersion, while we recover radial velocities for S/N$>$5 pix$^{-1}$.

Besides the models from \cite{Coelho2014}, we also tried PHOENIX stellar atmosphere models that provide a much higher spectral resolution of R=500,000, with a wide metallicity range $-4.0 <$ [Fe/H]$<+1.0$. We found no significant differences in the resulting dispersions and velocities, although we note that the PHOENIX model fits are notably worse (for reasons unknown) that those for the \cite{Coelho2014} models or Gaia-ESO empirical templates.

The resulting velocities and velocity dispersions are given in Table~\ref{table1}. The radial velocities listed are those obtained using the \cite{Coelho2014} models, while the dispersions are given for both the Gaia-ESO empirical templates and \cite{Coelho2014} models. While both the \cite{Coelho2014} models and Gaia-ESO stars give comparable quality fits for the velocity dispersions, we prefer those obtained from the Gaia-ESO stars since they should provide better representations of the Calcium Triplet lines (especially the line cores) than the \cite{Coelho2014} models which are calculated assuming local thermodynamic equilibrium.

\begin{table*}
\caption{Radial velocities, velocity dispersions, photometry and other relevant data for the DF2 GCs.}             
\label{table1}      
\centering                          
\begin{tabular}{l r c c c c c c c c}        
\hline\hline                 
GC & S/N & RV & $\sigma_{\rm gl}$ (Gaia-ESO) & $\sigma_{\rm gl}$ (Coelho) &  F606W$_0$ & F814W$_0$  & $r_{\rm eff}$ & $M/L_V$\\
 & (pix$^{-1}$) & (km s$^{-1}$) & (km s$^{-1}$) & (km s$^{-1}$) & AB(mag) & AB(mag) & (arcsec) &  ($M_\odot/L_\odot$)\\
\hline                        
   GC39   & 10.0 & $1798.2_{-1.5}^{+1.2}$& 6.1$_{-2.7}^{+2.0}$ &5.5$_{-2.1}^{+2.3}$& 22.37 $\pm$ 0.02 &  21.95 $\pm$ 0.02 & 0.056 $\pm$ 0.004 & 1.71 $\pm$ 1.32 &\\
   GC59   & 5.9 & $1782.8_{-3.4}^{+3.6}$& ... & ... &22.87 $\pm$ 0.03 & 22.31 $\pm$ 0.03 & 0.036 $\pm$ 0.002  & &\\
   GC71   & 6.8 & $1795.1_{-4.9}^{+3.8}$& ... & ... &22.65 $\pm$ 0.02 & 22.19 $\pm$ 0.03 & 0.047 $\pm$ 0.003  & &\\
   GC73   & 19.0 & $1797.7_{-1.2}^{+1.4}$&10.3$_{-1.7}^{+1.7}$ &10.9$_{-1.6}^{+2.2}$ & 21.53 $\pm$ 0.02  & 21.06 $\pm$ 0.02 &0.039 $\pm$ 0.002 & 1.57 $\pm$ 0.52&\\
   GC77   & 12.5 & $1799.9_{-1.1}^{+0.9}$&9.7$_{-2.0}^{+2.0}$ &10.1$_{-2.1}^{+2.3}$ & 22.10 $\pm$ 0.02  &  21.70 $\pm$ 0.02 &0.086 $\pm$ 0.003 & 5.18 $\pm$ 2.14 &\\
   GC85   & 10.3 & $1792.4_{-0.9}^{+0.7}$&7.6$_{-1.3}^{+1.5}$ &8.5$_{-1.4}^{+1.2}$ &22.37 $\pm$ 0.02   & 21.84 $\pm$ 0.02 &0.032 $\pm$ 0.002 & 1.55 $\pm$ 0.58 &\\
   GC91   & 7.5 & $1796.8_{-3.7}^{+4.6}$& ... & ... & 22.51 $\pm$ 0.02&  22.09 $\pm$ 0.03& 0.072 $\pm$ 0.002&  \\
   GC92   & 11.6 & $1779.8_{-1.3}^{+1.1}$& 7.1$_{-1.3}^{+2.3}$&8.6$_{-3.2}^{+1.8}$ & 22.13 $\pm$ 0.02 & 21.63 $\pm$ 0.02& 0.027 $\pm$ 0.001 & 0.90 $\pm$ 0.46\\
   GC98   & 6.1 & $1776.7_{-4.4}^{+5.7}$& ... & ... &22.76 $\pm$ 0.02 &22.29 $\pm$ 0.03   &0.028 $\pm$ 0.002 & & \\
   GC101  & 5.4 & $1836.3_{-3.9}^{+2.2}$& ... & ... & 22.86 $\pm$ 0.03&  22.38 $\pm$ 0.03 &0.034 $\pm$ 0.002 & & \\
\hline
 GC93   & ... & ... & ... & ... & 22.76 $\pm$ 0.02 & 22.29 $\pm$ 0.03& 0.028 $\pm$ 0.002 & ... \\
 GCNEW2 & ... & ... & ... & ... & 23.76 $\pm$ 0.04 & 23.25 $\pm$ 0.04& 0.044 $\pm$ 0.002& ... \\
 GCNEW3 & ... & ... & ... & ... & 22.58 $\pm$ 0.02 & 22.16 $\pm$ 0.03& 0.038 $\pm$ 0.002& ... \\
 GCNEW4 & ... & ... & ... & ... & 24.08 $\pm$ 0.04 & 23.51 $\pm$ 0.05 & 0.021 $\pm$ 0.005& ... \\
 GCNEW5 & ... & ... & ... & ... & 23.90 $\pm$ 0.04 & 23.43 $\pm$ 0.05& 0.081 $\pm$ 0.001& ... \\
GCNEW6\tablefootmark{a} & ... & ... & ... & ... & 23.69 $\pm$ 0.04 & 23.22 $\pm$ 0.04& 0.051 $\pm$ 0.002& ... \\
 GCNEW9 & ... & ... & ... & ... & 23.70 $\pm$ 0.04 & 23.32 $\pm$ 0.04& 0.134 $\pm$ 0.001& ... \\
\hline   
\end{tabular}
\tablefoot{
Upper part of table indicates GCs for which we obtained FLAMES spectra.
\\
\tablefoottext{a}{GCNEW6 has no spectroscopic confirmation.}
}
\end{table*}

\subsection{DF2 GC photometry and HST imaging}
\label{DF2GCPhotometry}

HST / Advanced Camera for Surveys (ACS) F606W magnitudes for the DF2 GCs have been published in \cite{vanDokkum2022_GCs}\footnote{F606W magnitudes are also presented in \cite{vanDokkum2018_GCs} and \cite{Shen2021_GCLF}, but we assume that those in \cite{vanDokkum2022_GCs} are the final version.} and \cite{Trujillo2019}. From \cite{vanDokkum2022_GCs} we have taken the F606W absolute magnitudes and converted them back to apparent magnitudes assuming the distance of 21.7 Mpc. Half-light radii ($r_h$) of the DF2 GCs have been published in \cite{vanDokkum2018_GCs} and \cite{Trujillo2019}. The absolute half-light radii from \cite{vanDokkum2018_GCs} were converted to apparent radii assuming their original distance of 19.0 Mpc \citep{vanDokkum2018_GCs}.

\begin{figure}
  \includegraphics[width=1.1\linewidth]{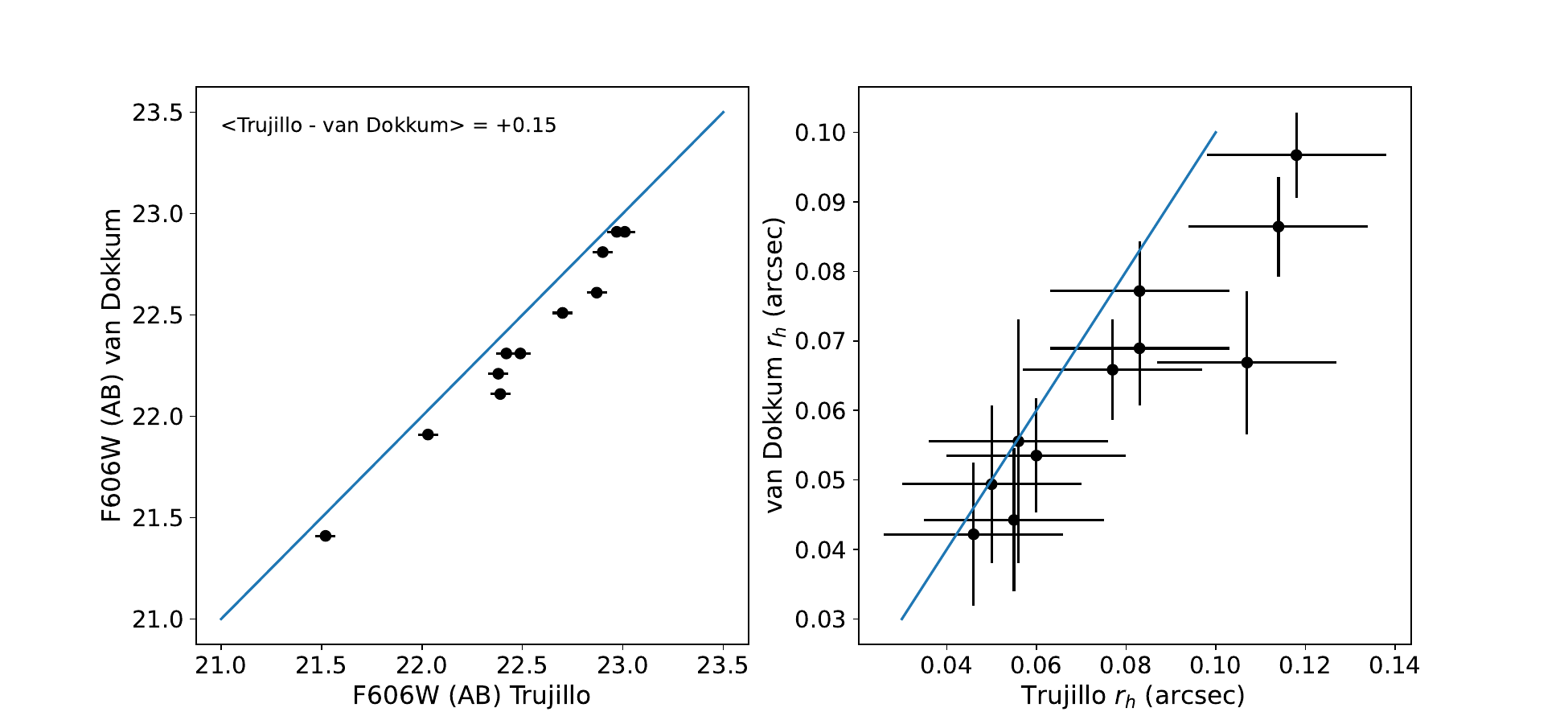}  \\

  \caption{Comparison of F606W magnitudes and half-light radii for DF2 GCs published by \cite{vanDokkum2022_GCs} and \cite{Trujillo2019}. Both the F606W magnitudes and sizes show important differences. The photometric uncertainties published by \cite{vanDokkum2022_GCs} are smaller than the symbol size.} \label{compare_trujillo_vdk}
\end{figure}

On inspection, we find that there are important differences between the apparent magnitudes and half-light radii measured for the GCs between these authors.  We show this in Fig.~\ref{compare_trujillo_vdk}. 
For the F606W magnitudes we find a mean difference of F606W(Trujillo) -- F606W(van Dokkum) of 0.15 mag. This is a significant offset, especially given the fact that the published uncertainties on these magnitudes is of order 0.02 mag for \cite{Trujillo2019} and $<0.015$ mag for \cite{vanDokkum2022_GCs}. In terms of half-light radii, for the smallest sizes ($r_h\sim0.05$"),  $r_h$ (Trujillo) is on average 10\% larger than $r_h$ (van Dokkum). For the largest GCs, this difference grows to $>20\%$.

The origin on these differences is not clear. For example, both sets of F606W magnitudes are measured in the AB system and corrected for extinction.
However, given these discrepancies, we have determined independent magnitudes and effective radii by analysing deep 40-orbit HST/ACS data from archive.
This is important not only for distance estimates with GCVD which requires accurate magnitudes, but also in order to determine the $M/L_V$ of the GCs which require accurate magnitudes, distances and sizes (Sect.~\ref{Masstolightratios}).

DF2 was observed with the HST ACS Wide-Field Channel (WFC) as part of the programmes 14644 and 15851 (PI: van Dokkum) The charge-transfer efficiency (CTE) corrected data were downloaded
from
MAST\footnote{\url[https://mast.stsci.edu/portal/Mashup/Clients/Mast/Portal.html[1]]} and \texttt{flc} files were used to build the final
drizzled mosaics (\texttt{drc}) with Astrodrizzle \citep{Gonzaga2012}.
The total exposure time is 40560s in F606W and 41762s in F814W.

In the following, we derive the profiles of the spectroscopically confirmed GCs of DF2 in order to be able to determine the parameters for performing accurate photometry and measuring their effective radii. 

\subsubsection{GC profiles} 
\begin{figure}
  \centering
  \subfloat{\includegraphics[width=0.8\linewidth]{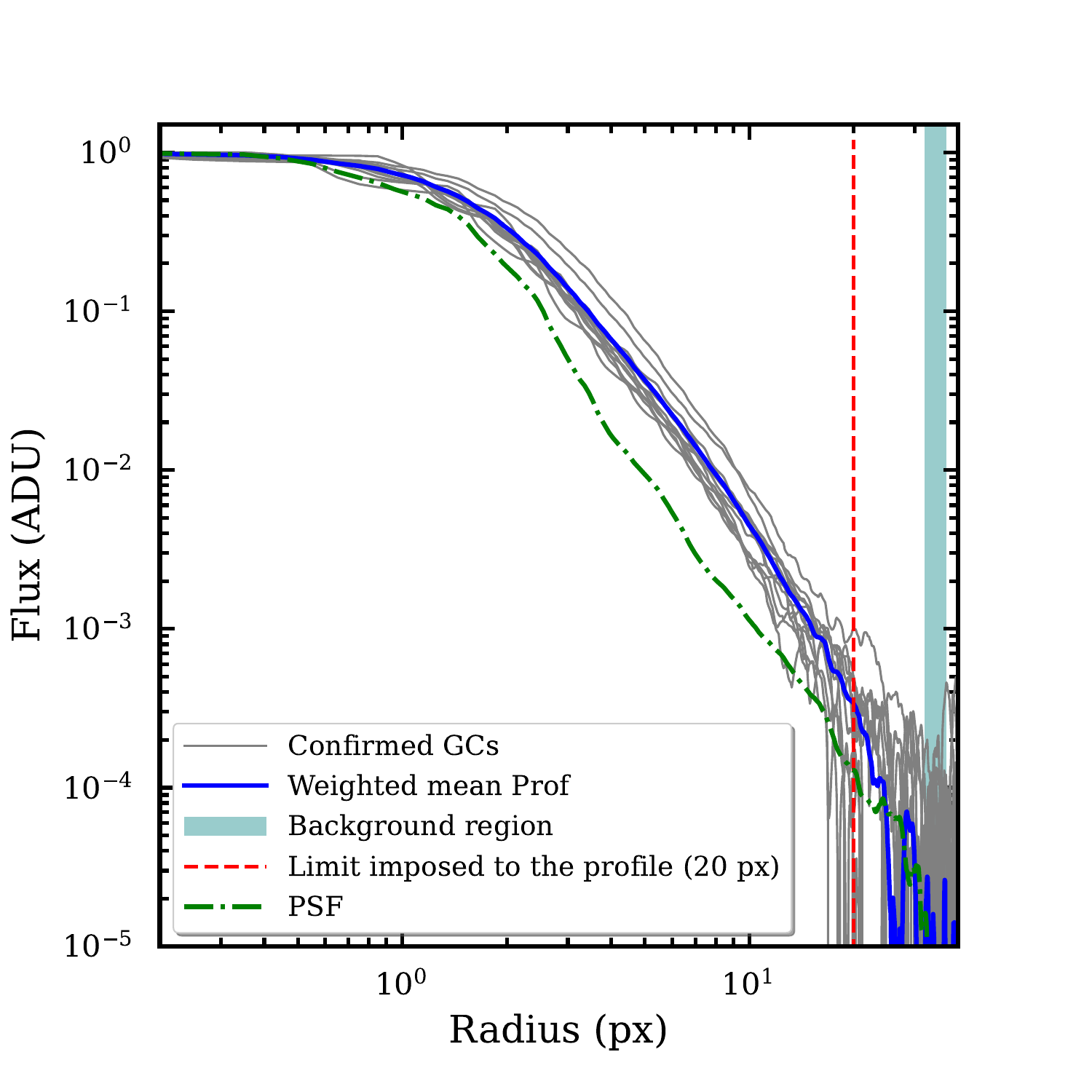} } \\
  \caption{GC profiles from HST/ACS F606W imaging for the 12 spectroscopically confirmed GCs from \cite{Trujillo2019} (grey), the signal-to-noise weighted mean profile of the GCs (blue line) and the mean profile of a star in the images (green). Additionally the vertical dashed red line indicates where the profile is cut and the light blue region indicates where the background has been estimated. The GCs are clearly more extended than the PSF.} 
  \label{Profiles}
\end{figure}

Light profiles of the spectroscopically confirmed DF2 GCs allow us to determine what parameters (aperture and flux corrections) are appropriate for performing accurate photometry of the GCs. The measurement of the half-light radii is also carried out on these profiles. The light profiles of the GCs were obtained on a large enough stamp (81x81 px) centred on the sources, to allow us to explore the profiles until they sink in the noise. On these stamps, we used the utility \texttt{centroid\_2dg} belonging to the \texttt{photutils} python package \citep{photutils2024}, to obtain the centroid of the sources. Then, the local background value was characterised using an annulus centred on the previously calculated centroid, and this background value was subtracted from every pixel of the stamp. For the selection of the background region, a variety of different radii and widths for the annuli were explored, looking for the annulus which minimises the RMS of the background region of the individual GC profiles. We adopted annuli of width 5 px with inner radii of 32 px. Once the background-subtracted stamps and the centroids of the profiles were determined, we computed the radial profiles using the \texttt{RadialProfile} utility (again, part of the \texttt{photutils} package). 
After some experimentation, we elected to use a mean GC profile, rather than use individual profiles, due to the fact that even with these deep data the profiles become quite noisy at larger radii, especially for the fainter GCs. 
To combine all the individual profiles and derive a representative profile, we computed the mean profile at each radius weighting by the signal to noise ratio of each of the profiles. Fig.~\ref{Profiles} shows, for the F606W imaging, the radial profiles for the 12 spectroscopically confirmed GCs from \citet{Trujillo2019} (grey), the SNR weighted mean profile (blue), the region where the background is estimated (light blue region) and the profile of the PSF (obtained from averaging different stars of the image). All the profiles have been normalised to 1 in the centre and, based on its signal-to-noise, cut at the same radii (20 px) before they become too noisy.

Having obtained the weighted mean profile, we derive from them the appropriate aperture to perform photometry. For this, we have estimated the FWHM of the profiles by finding the radius at which the flux is half of that at the centre. Also, we used the profiles to derive corrections to infinity:
    For this, we compared the amount of flux that is enclosed in apertures of $2\times$FWHM with the amount of total flux of the profile. Thus, this ratio gives us the correction that accounts for how much flux we are losing due to using finite apertures. Not only is this correction needed, but we also lose flux because we are unable to follow the profiles to infinity. They sink into the noise at some radius so we can only characterise them, in this case, up to approximately 20 px. This has to be taken into account and corrected which we did by analysing the slope in the outer regions of the PSF and extending our profiles with an exponential decay of this slope. Having completed these steps, the parameters obtained and used are listed in Table \ref{table2} and the final extinction-corrected F606W and F814W apparent magnitudes (AB system) are listed in Table \ref{table1}. Note that, for completeness, we also list photometry for GC93 from \cite{vanDokkum2018_DM} and also the GCs listed in \cite{Trujillo2019} which are not included in the present FLAMES observations.

\begin{table}[h]
\caption{Parameters for GC photometry}             
\centering   
\begin{tabular}{l c c c c} 
\hline\hline   
Filter & Aperture & Aperture & Profile corr. & Infinity corr.\\
 & (px) & (arcsec) & (mag) & (mag) \\
\hline 
F606W & 3.1 & 0.155 & 0.596 & 0.96$\pm$0.01\\
F814W & 3.1 & 0.155 & 0.574 & 0.95$\pm$0.01\\

\end{tabular}
\label{table2}      
\end{table}

To determine the photometric errors we modelled an artificial GC by circularising the mean GC profile (Fig. \ref{Profiles}). Then we injected this model in the images and compared the magnitudes that we inject with the magnitudes that we recover from our photometry. In order to obtain representative errors we injected sources in a region of two times the radius that encloses half of the GC population centered in the galaxy. The characterisation was performed between magnitudes 21--27, performing 15 steps of 6000 injections each. As a test of our adoption of a mean GC profile, we also derived photometry using individual profiles for the 5 brighter GCs for which we have velocity dispersions. The mean magnitudes for these five GCs using individual profiles is the same as when we use the mean profile. In detail, the individual magnitudes derived with individual profiles agree with the individual magnitudes derived using the mean profile to within 0.05 mag.

\subsubsection{Half-light radii}
\label{Half-lightradii}

We determined half-light radii ($r_{h}$) as follows: We used \texttt{pyimfit} (a python wrapper for \texttt{imfit}, \citealt{imfit2015}) to fit a function which approximates the intrinsic profile of GC, which we assumed to be a 2D modified King profile with $\alpha = 2$,  convolved with the PSF of the images. To model the PSF, we used a set of bright, non-saturated and non-contaminated stars.

The fits were performed on the F606W images in stamps of the sources with 81 pix $\times$ 81 pix. The background was estimated in the same region as for the photometry. The neighbouring sources were masked as best as possible, although the background is quite "granular" due to the partially resolved nature of the galaxy. Thus, the galaxy shows structure in its diffuse light, which makes the construction of the mask challenging. For the estimation of the errors, we provided \texttt{pyimfit} with an error map which is the standard deviation of the background region. The modified King profile parameters were chosen based on the typical values for Milky Way GCs and a range of possible distances to the galaxy. This way, the core radius ($r_c$) was set to vary between less than a pixel to a few pixels (physical values of one to a few pc) and the tidal radius ($r_t$) is imposed to be at least 6 pixels (typical range between 20 pc and 50 pc, although larger values are not unusual). The ellipticity was allowed to vary freely. We found that the ellipticity estimations are quite degenerate but the effective radius measurements are robust.\\

To summarize, the ranges for the parameters of the fits are the following: $0^{\circ}$ $\leq$ PA $\leq$ $180^{\circ}$; 0 $\leq$ Ellipticity $\leq$ 1.0; 0.16 px $\leq$ $r_c$ $\leq$ 2.5 px; $r_t$ $\geq$ 6px.
The half-light radii were then circularised using the expression $r_{h, circ} =  r_{h, SMA} * \sqrt{1 - ellipticity}$. The resulting $r_h$ of the GCs are listed in Table~\ref{table1}. For the clusters listed in the table, we find a median $r_{h} = 3.0\pm0.5$ pc adopting $d = 16.2$ Mpc as described in Sect.~\ref{Analysis}. These sizes compare very favourably with the median\footnote{We use the median here, since the size distribution of the Milky Way GCs is clearly non-gaussian with a strong tail to larger sizes.}  for the Milky Way GCs $r_{h} = 3.2\pm0.6$ pc, calculated from the 112 GCs in the catalogue of \cite{BaumgardtHilker2018}. 

\section{Analysis}
\label{Analysis}

 \begin{figure}
   \centering
   \includegraphics[width=\hsize]{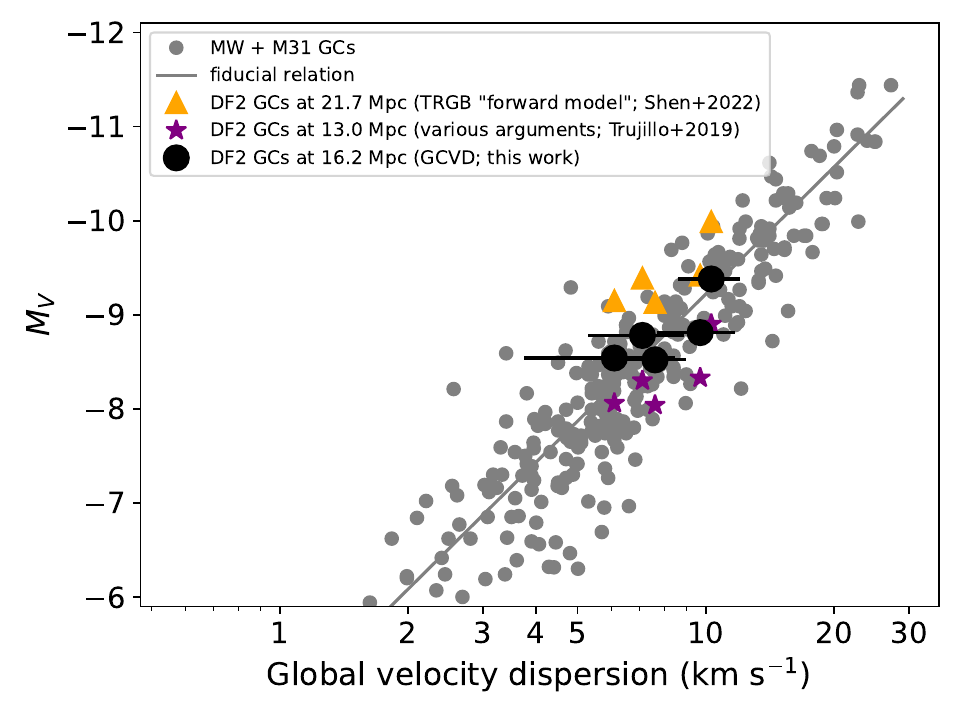}
      \caption{Relation between $M_V$ and the internal velocity dispersion for Milky Way and M31 GCs (grey circles). The grey line represents the fit to the relation as described in BFG24.
      The black points show the location of the DF2 GCs for our fit to this relation giving $d=16.2$ Mpc. Yellow triangles show the location of the DF2 GCs assuming a further distance ($d=21.7$ Mpc; \cite{Shen2023}), purple stars show the the DF2 GCs assuming $d=13$ Mpc \citep{Trujillo2019}. For the shorter and further distances the DF2 GCs lie systematically off the relation defined by the Milky Way and M31 data.
              }
         \label{GCVD}
   \end{figure}

\subsection{Distance to DF2}

To obtain the distance to NGC1052-DF2, we used the "globular cluster velocity dispersion" approach (BFG24). 
GCVD relies on the fact that GCs in the Local Group obey a common $M_V - \sigma_{\rm gl}$ relation. 

BFG24 showed that the Milky Way and M31 GCs obey the following relation:

\begin{equation}
    M_V = \beta_0 + \beta_1{\rm log_{10}}(\sigma_{\rm{gl}}),
    \label{eq1}
\end{equation}

\noindent where $\beta_0$ is the zeropoint and $\beta_1$ the slope of the relation.

If we assume that the DF2 GCs also obey this relation, then we may fix the slope ($\beta_1$) to the fiducial value given in BFG24 ($\beta_1=-4.73^{+0.05}_{-0.05}$). A fit to the above relation (i.e., corresponding to a simple shift in zeropoint) with our measured velocity dispersions then yields $\beta_0 \rm{(DF2)}$. The distance modulus to DF2 can then be calculated from:

\begin{equation}
    (m - M)_0 \rm{(DF2)} =  \beta_0 \rm{(DF2)} - \beta_0 \rm{(fiducial)},
    \label{eq2}
\end{equation}

\noindent with $\beta_0 \rm{(fiducial)} = -4.49^{+0.04}_{-0.04}$ (BFG24).\\

The GCVD is based on Johnson V-band (Vega) magnitudes for Milky Way and M31 GCs. Since the magnitudes for the DF2 GC were obtained using the F606W ACS/WFC filter in the AB system, we obtained a conversion between the systems using the e-MILES SSP models. With the MILES webtools\footnote{http://research.iac.es/proyecto/miles/pages/webtools.php} we convolved a set of 15 SSP models for 3 ages (8, 10, 12 Gyr) and 5 metallicities ([M/H] = --0.66, --0.96, --1.26,--1.49 and --1.79) with the F606W and V filters, and applied the appropriate zeropoints. The choice of model ages and metallicities was informed by the stellar population results from Paper II (i.e., old, metal-poor). We then obtained and applied the mean correction to the F606W magnitudes: V(Johnson,Vega) = F606W(AB) + $0.14\pm0.01$, where the uncertainty on the correction is the standard deviation on the mean differences for the 15 models. This error is propagated through the analysis. 

With V-band apparent magnitudes and velocity dispersions in hand, we used an Monte Carlo Markoc Chain (MCMC) ensemble sampler (\textsc{emcee}; \citealt{Foreman-Mackey2013}) to determine $\beta_0$ and its uncertainties. We adopt a Gaussian likelihood function and include uncertainties in both $V$  and $\sigma_{\rm{gl}}$.  We chose uniform priors in the range $-10 < \beta_0 < 31$. The large range in $\beta_0$ encompasses distances from the Milky Way out to approximately 100 Mpc. We used 25 walkers and 10,000 chains. The uncertainties on the parameters are given as the 16th and 84th percentiles of the resulting distributions.  

From the MCMC for the DF2 GCs we find $\beta_0 \rm{(DF2)}=26.56\pm0.16$. Since the $V$ magnitudes are already corrected for extinction, from this we obtain directly a distance modulus $(m - M)_0 \rm{(DF2)}=31.05\pm0.17$, or a distance of $d = 16.2\pm1.3$ (stat.) $\pm1.7$ (sys.) Mpc. This distance is obtained using a combination of the photometry derived in Sect.~\ref{DF2GCPhotometry} and the velocity dispersions using the Gaia-ESO stellar templates (Sect.~\ref{Measurementofradialvelocitiesandvelocitydispersions}).
Our estimation of the uncertainties is discussed in Sect.~\ref{Statisticalandsystematicuncertainties}. 

Our fits for the distance are shown in Fig.~\ref{GCVD}, where we compare the DF2 GCs --shifted by a distance of 16.2 Mpc -- to the fiducial relation defined by the Milky Way and M31 GCs. We also show in the figure the positions of the DF2 GCs assuming distances of 21.7 Mpc \citep{Shen2023} and 13.0 Mpc \citep{Trujillo2019}. For the further distance of 21.7 Mpc the DF2 GCs lie systematically above the relation by an average $\sim0.6$ magnitudes, whereas for the closer 13.0 Mpc distance the GCs lie below the relation by $\sim0.4$ mag. The slope of the relation is not well constrained due to the small sample size, however, visually the DF2 GCs seem to follow that of the combined Milky Way and M31 relation quite well.

\subsection{Statistical and systematic uncertainties}
\label{Statisticalandsystematicuncertainties}

The statistical uncertainty on the distance is determined from the MCMC when fitting for the zeropoint of the $M_V - \sigma_{\rm gl}$ relation, taking into account the uncertainty in the zeropoint of the fiducial relation (BFG24). The magnitude of the statistical uncertainty is affected by a combination of the sample size, the range of $\sigma_{\rm gl}$ in the data (a larger range in velocity dispersion better samples the relation) and the uncertainties in $\sigma_{\rm gl}$ and the V-band apparent magnitudes. In the present case, with velocity dispersions for 5 DF2 GCs, we obtain a statistical uncertainty on the distance of $\sim8$\%.

\begin{table}[h]
\caption{Distances derived using different velocity templates and GC photometry}             
\centering   
\begin{tabular}{l c } 
\hline\hline   
Templates / Photometry & Distance \\
 & (Mpc) \\
\hline 
Gaia-ESO / this paper & 16.2$\pm1.3$\\
Gaia-ESO / \cite{vanDokkum2022_GCs} & 15.6$\pm1.2$\\
\cite{Coelho2014} / this paper & 16.7$\pm2.1$\\
\cite{Coelho2014} / \cite{vanDokkum2022_GCs} & 16.1$\pm2.1$\\
\hline
\end{tabular}
\label{table3}      
\end{table}

Characterisation of the systematic uncertainties includes accounting for the variations in $\sigma_{\rm gl}$ obtained from different spectral templates, differences in the zeropoints of the GC photometry, variations in the intrinsic sizes of the GCs, aperture and fibre centring effects and uncertainties in the intrinsic $M/L_V$ of the DF2 GCs. This final point -- the mass to light ratios of the clusters -- is  discussed in detail in Sect.~\ref{Masstolightratios}.

\subsubsection{Systematic error due to intrinsic size variations}
\label{Systematicerrorduetosizevariations}

An additional source of systematic uncertainty on the distances is related to the intrinsic sizes of the GCs. In the Milky Way there is a tendency for larger GCs (in terms of $r_h$) to have lower $\sigma_{\rm gl}$ for a given $M_V$ \citep{McLaughlin2005}.  As discussed in Sect.~\ref{Half-lightradii}, we obtain a median $r_{h} = 3.0\pm0.5$ pc when adopting $d = 16.2$ Mpc, while assuming $d = 21.7$ Mpc \citep{Shen2023}, we obtain a median $r_{h} = 4.0\pm0.7$ pc. We investigated this potential systematic by only selecting larger Milky Way and M31 GCs such that they too have a median $r_{h} = 4$ pc. Re-running GCVD based only on this subset of clusters yields a GCVD distance of $17.2\pm1.6$ Mpc (stat.).  I.e., if we underestimate the cluster size, the distance may also be underestimated - in this case by about $\sim6\%$.~\cite{vanDokkum2018_GCs} find a mean size of $r_{h} = 8.4$ pc for the DF2 GCs (assuming the distance of \citealt{Shen2023}).
We find no evidence for very large  GCs even assuming $d = 21.7$ Mpc.  However, although we have few very large clusters in our combined Milky Way and M31 sample (6 GCs with  $r_{h} > 8$ pc, only two of which have $\sigma_{\rm gl} > 4$ km/s which is the useful region for GCVD distances as discussed in \cite{Beasley2024}), we estimate that by selecting GCs only with $<r_{h}> = 8.4$ pc would shift the inferred distance to about 18.0 Mpc (with large uncertainties since we do not fit the distance reliably with 2 datapoints). In summary, based on our new sizes, potential size differences between the GCs defined in the fiducial GCVD relation and those in DF2 may affect our distance estimates by at most $\sim10\%$.

\subsubsection{Aperture corrections and FLAMES fibre centring}
\label{AperturecorrectionsandFLAMESfibrecentring}

The velocity dispersions we measure for the DF2 GCs are affected by the finite aperture of the FLAMES fibres and also by their positioning error, and we account this in our analysis (see also \cite{Hilker2007} and \cite{Mieske2008}). The effect of finite apertures has a tendency to lead to overestimates of the true velocity dispersion since light is preferentially lost from the outer regions of the clusters, and GCs generally exhibit declining velocity dispersion profiles with radius. In addition, there is an uncertainty associated with the positioning accuracy of the FLAMES fibres themselves. This has the tendency to lead to an underestimation of the true velocity dispersion, since light is preferentially lost from the central regions of the cluster due to imperfect centring. I.e., the effects of aperture and fibre centring work in opposition to each other. 

To assess the importance of these effects we ran a series of simulations on artificial GCs as described in detail in Appendix~\ref{appendix_B}. We find that the combined effects lead to a general underestimate of the true velocity dispersion by $\sim1\%$; for distances $>10$ Mpc, the uncertainty in the FLAMES fibre centring dominates any light losses due to aperture. However, the correction is sensitive to seeing, distance and intrinsic GC size (see Fig. \ref{appendix_Fig3} in Appendix~\ref{appendix_B}). Given this, we add this as a typical uncertainty to our total systematic error budget (which amounts to $\sim0.15$ Mpc) rather than attempt to correct the distance directly. 

\subsubsection{Effect of different spectral templates and photometry}
\label{Affectofdifferentspectraltemplatesandphotometry}

We used two different sets of spectral templates (Gaia-ESO and the \citealt{Coelho2014} models) and can use two different sets of apparent magnitudes (those derived by us, and those of \citealt{vanDokkum2022_GCs}) to have four different distance determinations using GCVD. While we favour the distance returned using the Gaia-ESO templates and our magnitudes, as justified previously, the other three distance estimates can give us an idea of the range of values due to the different template/model combinations. These different distances are shown in Table~\ref{table3}.

These distances range from 15.6--16.7 Mpc, and 
the mean of these distances coincides with our "favoured" distance of 16.2 Mpc. We note that the distances determined when using the photometry of \cite{vanDokkum2022_GCs} are 0.6 Mpc closer than when using our photometry for a given set of templates. This arises because the magnitudes from \cite{vanDokkum2022_GCs} are systematically brighter than ours by an average 0.18 mag. The standard deviation of the four GCVD distances is 0.39 Mpc, and we include this in the systematic uncertainty.

Combining all the above uncertainties gives our final systematic uncertainty on the distance of 1.7 Mpc, or $\sim11\%$ at $d=16.2$ Mpc.

\subsection{Mass to light ratios}
\label{Masstolightratios}

The mass-to-light ratios of the DF2 GCs  bring important constraints on their nature.  
With the absolute V-band magnitudes and half-light radii in hand, we can determine GC masses and $M/L_V$.

We obtained virial masses for the DF2 GCs using the following equation defined in terms of 2-D (projected) observables (e.g., \citealt{Larsen2002}; \citealt{Gvozdenko2024}): 

\begin{equation}
M_{\rm vir} = 10 \frac{\sigma_{\rm {gl}}^2 r_{\rm {h}}}{G}
\label{eq3}
\end{equation}

The prefactor of 10 arises from the
constant of 2.5 given in \cite{Spitzer1987}, multiplied by a conversion from 3-D to 2-D quantities where $\sigma^2_{\rm 3D} = 3 \sigma^2_x$ and $r_{\rm {hm, 3D}} = 4/3 r_{\rm {h, 2D}}$. Here, $r_{\rm {hm, 3D}}$ is the 3-D half-mass radius and we make the assumption that mass follows light in the GCs. The $M/L_V$ ratios determined for the DF2 GCs are given in Table~\ref{table1}, where we give the results using the Gaia-ESO templates, assuming a distance of 16.2 Mpc. The results for the \cite{Coelho2014} are similar, but on average yield $\sim$10\% larger $M/L_V$. 

\cite{Baumgardt2020} give $M/L_V$ for 150 Milky Way GCs determined from stellar velocity profile fits to a large suite of N-body simulations. In order to compare directly with our results, we re-derived $M/L_V$ for their sample using Eq.~\ref{eq3}, i.e. using the global velocity dispersion and half light radius only. We converted central velocity dispersions given in \cite{Baumgardt2019} to global velocity dispersions (i.e., velocity dispersions to "infinite" radius)  using the following formula from BFG24:

\begin{equation}
\sigma_{\rm gl} = 0.085(\pm0.064) + 0.752(\pm0.009)\sigma_{\rm cen}
\label{eq4}
\end{equation}

\noindent where $\sigma_{\rm cen}$ is the central velocity dispersion. 

We then selected Milky Way GCs with $E(B-V) <$ 0.3 and photometric uncertainties $<$ 0.1 mag. This effectively excludes all the fainter clusters with $M_V > -5$ and many of the more metal-rich clusters which are affected by appreciable extinction. Using Eq.~\ref{eq3}, for the reduced sample (56 clusters), we obtain a mean $M/L_V = 1.77\pm0.10~M_\odot/L_\odot$. Using the $M/L_V$ given in \cite{Baumgardt2020} for this sample, we obtain a mean value $M/L_V = 1.78\pm0.07~M_\odot/L_\odot$. So, reassuringly, we find good agreement between the $M/L_V$ derived by \cite{Baumgardt2020} from resolved data, and the $M/L_V$ from the virial formula alone. 

\begin{figure}
   \centering
    \includegraphics[width=\hsize]{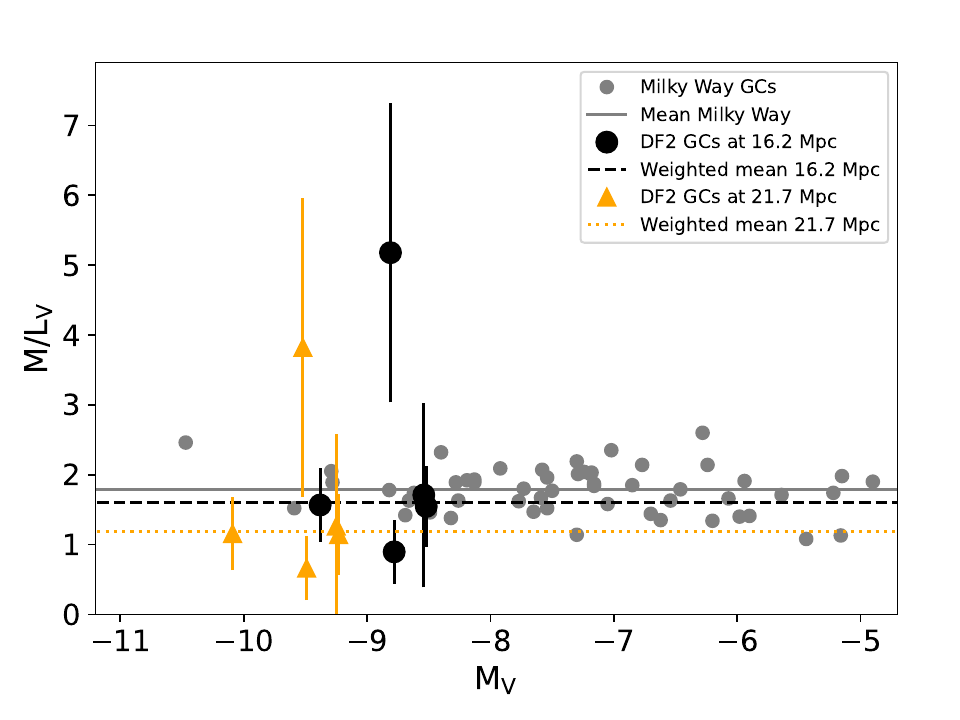}
   \caption{$M/L_V$ of the DF2 clusters compared to  Milky Way GCs (grey) for distances of 16.2 Mpc (black circles) and 21.7 Mpc (orange triangles). In general, the DF2 GCs fall below the Milky Way sample if the larger distance is assumed. The most luminous Milky Way  GC with $M/L_V\sim$2.5 is $\omega$ Cen (NGC~5139). }
              \label{ML}%
    \end{figure}

The $M/L_V$ ratios of the DF2 GCs are compared to the Milky Way GC sample in Fig.~\ref{ML}. For $d=16.2 $ Mpc, we obtain a weighted mean for the DF2 GCs, $M/L_V = 1.61\pm0.44~M_\odot/L_\odot$ (median $M/L_V = 1.57\pm0.55~M_\odot/L_\odot$). These mass-to-light ratios are in good agreement with Milky Way sample and is consistent with the finding that the mass function of Local Group GCs and massive clusters appears to be largely invariant with environment, metallicity, mass, density and velocity dispersion \citep{Baumgardt2023_MF}.

%

   \begin{figure}
   \centering
    \includegraphics[width=\hsize]{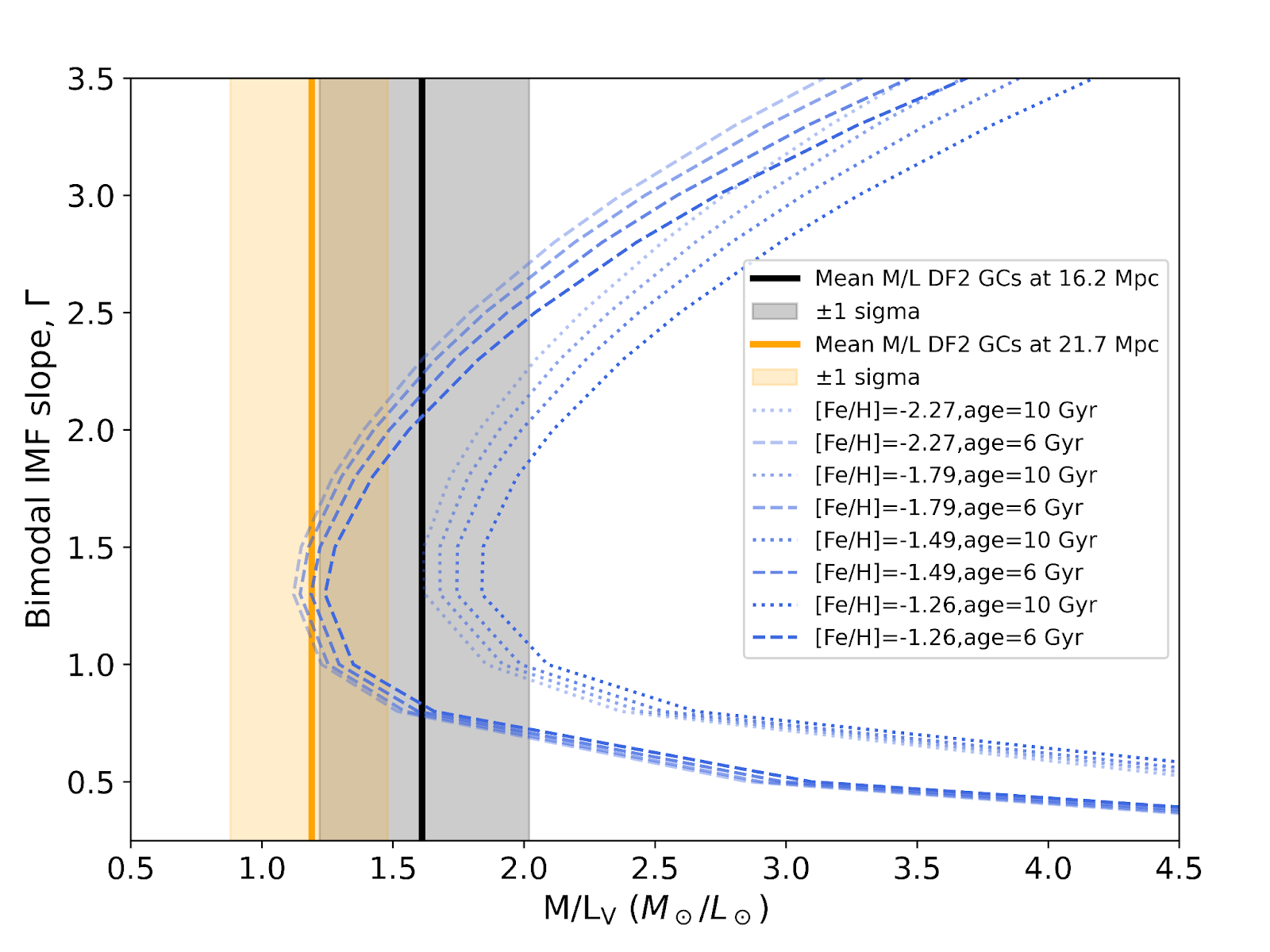}
   \caption{Weighted mean $M/L_V$ of the DF2 GCs for distances of 16.2 Mpc (black) and 21.7 Mpc (orange) compared to expectations from the MILES models \citep{Vazdekis2010} for a bimodal IMF with a range of slopes \citep{Vazdekis1996}. $\Gamma=1.3$ corresponds to a Kroupa IMF \citep{Kroupa2001}. A distance of 16.2 Mpc is consistent with the models for 10 Gyr, in agreement with the DF2 GC ages from MUSE data found in Paper II. A distance of 21.7 Mpc suggests a lower $M/L_V$ which cannot be reproduced for old ages with this parametrisation.
   }
              \label{ML_mod}%
    \end{figure}

On the other hand, adopting the distance of 21.7 Mpc, we find for the DF2 GCs $M/L_V = 1.19\pm0.33~M_\odot/L_\odot$ (median $M/L_V = 1.16\pm0.41~M_\odot/L_\odot$). I.e.,  adopting the further distance gives $M/L_V$ for the DF2 GCs which are notably lower than that of the Milky Way clusters. 

Exploring this some more, in Fig.~\ref{ML_mod} we compare the derived $M/L_V$ for the DF2 GCs to expectations from the MILES stellar population models \citep{Vazdekis2010} assuming the bimodal initial mass function (IMF) parametrisation defined in \cite{Vazdekis1996}. The figure shows the mean $M/L_V$ assuming distances of 16.2 and 21.7 Mpc for the DF2 GCs, for a range values for the bimodal IMF slope ($\Gamma_{\rm b}=0.3-3.5$), where $\Gamma_{\rm b}=1.3$ is very similar to the \cite{Kroupa2001} IMF. For $d=16.2$ Mpc, the DF2 GCs are consistent with a "Kroupa-like" IMF for ages of order 10 Gyr. The further distance suggests ages closer to $\sim6$ Gyr. Important additional constraints come from the stellar populations of the DF2 GCs (\citealt{vanDokkum2018_DM}; \citealt{Fensch2019}). In particular, in Paper II we derived ages and metallicities for 4 of the 5 individual GCs for which we have obtained velocity dispersions. The mean age for GC73, GC77, GC85 and GC92 is $10.1\pm0.8$ Gyr, with mean metallicity [M/H] = $-1.44\pm0.21$\footnote{GC39 is outside the MUSE footprint in Paper II, however its colours suggest close similarity to the other GCs. I.e., old and metal-poor.}. As can be seen, at $d=16.2$ Mpc the GCs are in good agreement with the models. In contrast, the lower $M/L_V$ implied for the GCs by assuming $d=21.7$ Mpc distance is inconsistent with the model predictions for all modelled IMF slopes and metallicities.

It is interesting to note that the bimodal IMF exhibits a minimum in $M/L$ at $\Gamma_{\rm b}=1.3$ (the "Kroupa-like" IMF) for intermediate to old stellar populations. This can be understood since for $\Gamma_{\rm b} > $ 1.3 -- "bottom heavy" IMFs -- the stellar population is dominated by low-mass stars which contribute relatively more mass than light. In contrast, for $\Gamma_{\rm b} < $ 1.3 -- "top heavy" IMFs -- in intermediate and old populations the massive stars have already died and left dark remnants which contribute mass but no light (see additional discussions in e.g., \citealt{Cappellari2012}, \citealt{Ferreras2013} and \citealt{FerreMateu2013}).

While the IMF in galaxies is not expected to be strongly modified over time via dynamical effects, this is likely not the case for GCs. Indeed, the present day stellar mass function (MF) of Milky Way GCs is probably different from their IMF at birth (e.g., \citealt{Vesperini1997}; \citealt{Baumgardt2003}; \citealt{Baumgardt2023_MF}). However, given the age and metallicity constraints for the DF2 GCs, for the distance of $d=21.7$ Mpc to be correct the DF2 GCs would likely be required to have a mass function significantly more depleted in low mass stars (and remnants) than the Milky Way GCs. 

\section{Discussion and Conclusions}
\label{Discussion}

\begin{figure}
   \centering
   \includegraphics[width=\hsize]{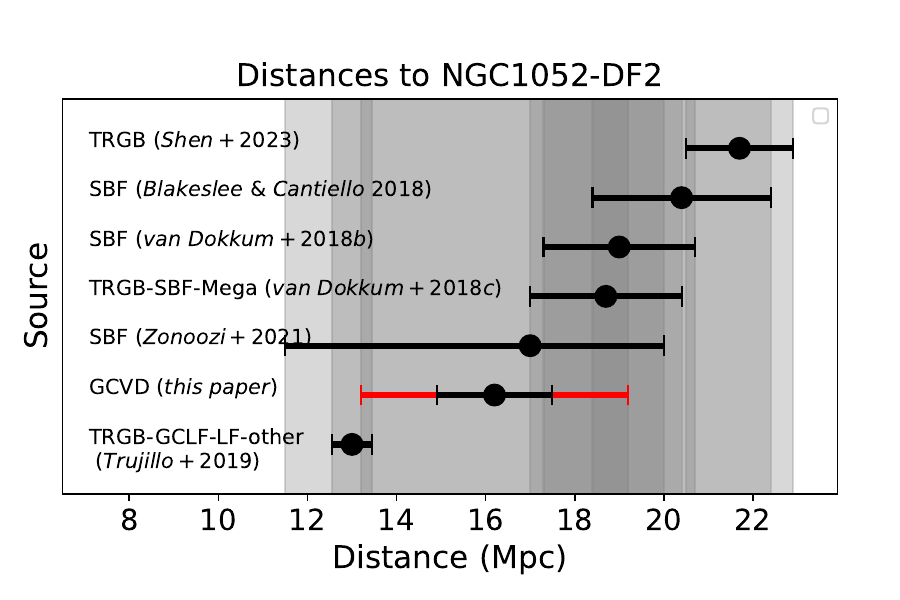}
      \caption{Distance estimates to NGC1052-DF2 from the literature compared to our determination. Black horizontal bars represent the statistical uncertainty from each source. In the case of our GCVD distance, we also show the estimated systematic uncertainty in red. Grey columns indicate the degree of overlap between the distances, where darker shades indicate more overlap. For the \cite{Shen2023} TRGB measurement no uncertainty is given, therefore we assume a statistical uncertainty $\pm1.2$~Mpc from \cite{Shen2021_TRGB_distance}.
              }
         \label{distances}
   \end{figure}

By assuming that the NGC1052-DF2 GCs obey the same $M_V - \sigma_{\rm gl}$ relation as Milky Way and M31 GCs, we use the "GCVD" technique (BFG24) to obtain a distance to NGC1052-DF2 of $d=16.2\pm1.3$ (stat.) $\pm1.7$ (sys.) Mpc. This places the galaxy somewhat closer to us than suggested by some other studies in the literature. A comparison of our distance to literature values is shown in Fig.~\ref{distances}.

Literature distance estimates for DF2 range from 11.5 Mpc \citep{Zonoozi2021} to 22.9 Mpc \citep{Shen2023} (including the 1$\sigma$ uncertainties). In the figure we have not included the additional distance constraints from the planetary nebulae (PNe) in DF2 from \cite{Fensch2019}, who find distances of 13 Mpc and 
20 Mpc equally likely based on the three identified PNe.

If we only consider the statistical uncertainties, our GCVD measurement is in agreement with the SBF distance of \cite{vanDokkum2018_DM}, the "TRGB-SBF-Megamaser" distance of \cite{vanDokkum2018_distance} and the SBF distance of \cite{Zonoozi2021}. We do, however, find disagreement with the TRGB distance of \cite{Shen2023} at the $\sim 3\sigma$ level, and disagreement with the shorter distance of \cite{Trujillo2019} at $\sim 2\sigma$. We note that the \cite{Trujillo2019} estimate is consistent with the SBF distance of \cite{Zonoozi2021}, but not with the rest of the published distances.
Inclusion of our systematic uncertainties brings marginal agreement between our distance and the SBF distance of \cite{Blakeslee2018} and with \cite{Trujillo2019}, but not with \cite{Shen2023}.  It is worth pointing out that the distance from \cite{Trujillo2019} is an average of five distance indicators (the GCLF, GC sizes, TRGB, a luminosity function comparison with DD044 and SBF). The SBF value of \cite{Trujillo2019}, $d$=14.7$\pm$1.7, which gives the furthest distance of the five methods, is in agreement with GCVD and is only marginally inconsistent with the TRGB-SBF-Megamaser distance.  We also note that the 21.7 Mpc TRGB distance of \cite{Shen2023} is also in marginal disagreement with the TRGB-SBF-Megamaser distance from \cite{vanDokkum2018_distance}.

Inspection of Fig.\ref{distances} suggests that the systematic uncertainties on the distances may be underestimated quite significantly (when considered at all).  
For example, the SBF analysis from \cite{Zonoozi2021} considers a range of star formation histories and non-canonical IMFs, which lead to a range of distances significantly larger than the uncertainties given for the other SBF (or TRGB) estimates. Systematic uncertainties in the SBF approach for the specific case of NGC1052-DF2 has been extensively discussed in \cite{Trujillo2019}, and in more general contexts in \cite{Greco2021}, \cite{Zonoozi2021} and \cite{Pablo2024}.

The TRGB is a well established method for measuring distances to galaxies. However, obtaining a reliable TRGB detection for distant, faint systems is challenging even for HST, which becomes limited by both crowding and photometric depth. This is especially true for low-mass systems like DF2. 
\cite{Shen2021_TRGB_distance} employ a "forward model" to identify the TRGB, rather than clearly identify the tip itself. So, while the TRGB technique in general is regarded as a powerful distance indicator, there is reason for some caution about the NGC1052-DF2 distance given that HST is on the edge of its technical capabilities.

On the other hand, it is appropriate to remark on the robustness of the GCVD distance to NGC1052-DF2.
It goes without saying  that obtaining both precise and accurate distances to distant, low-mass galaxies like NGC1052-DF2 is not straightforward. One particular issue we were very aware of during our analysis was possible human bias. We (the authors) were generally cognisant of the pre-existing distance estimates (and their possible implications) for NGC1052-DF2, and there are steps in any analysis which can be unconsciously influenced by such foreknowledge. We tried our best to mitigate this via duplicating analysis roles and comparing results. In addition, the final measurement of the GC velocity dispersions was performed by one of us (Fahrion) and kept separate from the fits to the distance (Beasley), while the GC photometry was performed independently (Guerra, Montes).

In terms of the GCVD absolute calibration, GCVD distances are anchored to Gaia DR2/3 distances of Milky Way GCs. BFG24 have shown that the technique works well for the Local Group galaxies they considered, and also for the relatively nearby giant elliptical NGC~5128 ($d=3.89\pm0.16$ Mpc). The GCVD distance to M31 is one of the more precise (and perhaps accurate) distances in the literature ($d=798\pm28$ kpc). In the case of relatively GC-poor Local Group dwarfs, GCVD returns typical statistical uncertainties of $\sim13\%$. However, we acknowledge that the present study is the furthest that the technique has been applied to date.

Some of the systematic uncertainties in GCVD are discussed in Sect.~\ref{Statisticalandsystematicuncertainties} and also in BFG24. We showed in Sect.~\ref{Masstolightratios} that for $d=16.2$ Mpc, the $V$-band mass to light ratios of the DF2 GCs (mean $M/L_V = 1.61\pm0.44~M_\odot/L_\odot$) are consistent with those of the Milky Way clusters (mean $M/L_V = 1.77\pm0.10~M_\odot/L_\odot$). 
Adopting the furthest literature distance \citep{Shen2023} yields lower values (mean $M/L_V = 1.19\pm0.33~M_\odot/L_\odot$) than are seen 
in Milky Way clusters or predicted by stellar population models for all reasonable IMFs (see Fig.~\ref{ML_mod}). The implicit assumption in GCVD is that the target GCs in question have similar $M/L_V$ to the Milky Way and M31 GCs. If this is not the case, then the distance estimates will be affected in the sense that for systematically lower (higher) $M/L_V$, distances will be underestimated (overestimated). In Paper II we determined the ages and metallicities of individual DF2 GCs and find that they are old and metal-poor. This is consistent with what was found previously using binned cluster spectra (\citealt{vanDokkum2018_DM}; \citealt{Fensch2019}). Taking into account their old ages, perhaps the only remaining way to lower $M/L_V$ in the DF2 GCs is to posit that their present-day stellar mass functions (MF) differ significantly from those seen in the Milky Way and M31 and, by inference from our results in BFG2024, from GCs in the Local Group generally.

We know that the present day stellar MFs of the Milky Way GCs are likely different from their IMFs (\citealt{Vesperini1997}; \citealt{Takahashi2000}; \citealt{Baumgardt2023_MF}). The
stellar MF of a GC is expected to be modified as consequence of mass segregation, whereby more massive stars sink into
the cluster centre and low-mass stars move outwards to regions
where they are preferentially lost to external tidal fields.
For example, the N-body simulations of \cite{Baumgardt2003} indicate that this process
primarily leads to a decrease of the GC $M/L$ ratio
since low-mass stars ($M_*<0.5$ M$_\odot$) contribute more mass than light than do more massive stars.
\cite{Baumgardt2003} found a maximum decrease of  the $M/L$ ratio of about $\sim$0.6 due to this mechanism.
However, it is not clear whether this process could explain the low $M/L_V$ ratios of the DF2 GCs, assuming a distance of 21.7 Mpc, since the tidal field of DF2 is presumably significantly weaker than that of the Milky Way due to its low DM fraction (see e.g., \citealt{Miholics2016} and also the discussion in \citealt{vanDokkum2018_GCs}). If anything -- due to the weaker tidal field of NGC1052-DF2 -- the GCs might be expected to have higher $M/L_V$ ratios than their Milky Way counterparts based on this argument.

In any case, even though the mechanism for lowering the $M/L_V$ of the GCs of NGC1052-DF2 might not be immediately obvious, the possibility remains that these GCs do differ from the Milky Way clusters in that they might have a "dwarf depleted" or "bottom-light" MF. A definitive answer to this, and to the broader issue of the precise distance to NGC1052-DF2 may have to wait for a clear detection of the tip of the RGB via JWST.

\begin{acknowledgements}
      We thank Alejandro Vazdekis, Nacho Mart\'in and Ignacio Trujillo for useful discussions. This project has received funding from the European Union’s Horizon Europe research and innovation programme under the Marie Sk\l{}odowska-Curie grant agreement No 101103830.
      KF acknowledges support through the ESA research fellowship programme.
      MM acknowledges support from the project RYC2022-036949-I financed by the MICIU/AEI/10.13039/501100011033 and by FSE+. 
      MAB, SGA  acknowledge support from the Spanish Ministry of Science and Innovation (MICIU) grant number PID2022-140869NB-I00.
      This work made use of Astropy:\footnote{\url{http://www.astropy.org}} a community-developed core Python package and an ecosystem of tools and resources for astronomy \citep{astropy2013, astropy2018, astropy2022}. Also we made extensive use of Numpy \citep{Numpy} and Scipy \citep{Scipy}. We warmly thank the ESO support staff for their help during the programme preparation and execution.
      Based on observations collected at the European Southern Observatory under ESO programme 110.23P4.001. This research is also partly based on observations made with the NASA/ESA {\it Hubble} Space Telescope obtained from the Space Telescope Science Institute, which is operated by the Association of Universities for Research in Astronomy, Inc., under NASA contract NAS 5–26555. These observations are associated with programmes 14644 and 15851. 
\end{acknowledgements}

%
%
\bibliographystyle{aa} 
\bibliography{references}

\appendix
\begin{appendix} 
\onecolumn

\section{PPXF fits to individual globular clusters}
\label{appendix_A}

\begin{figure}[h!]
   \includegraphics[width=\hsize]{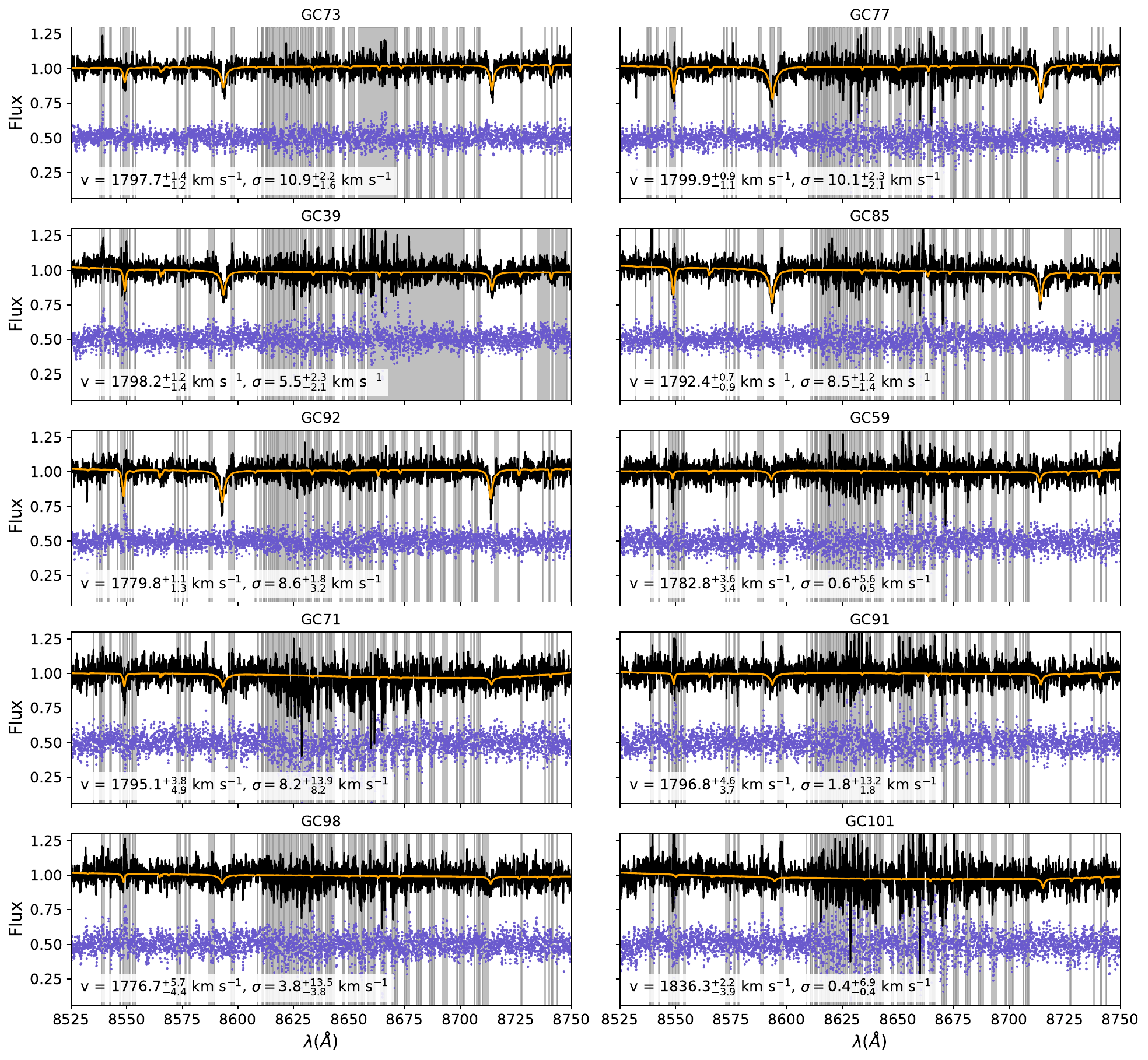}
      \caption{\textsc{pPXF} fits for the GC sample. GIRAFFE spectra are shown in black, the optimal model template in orange and residuals of the fit are shown underneath in purple. The greyed out areas are regions of strong sky residuals which are masked out in the fitting procedure. Radial velocities and velocity dispersions are shown for each corresponding GC. Velocity dispersion values less than $\sim4$ km/s are below the resolution limit of our analysis and therefore unreliable.
              }
         \label{appendix_Fig1}
   \end{figure}

\begin{figure}[h!]
   \includegraphics[width=\hsize]{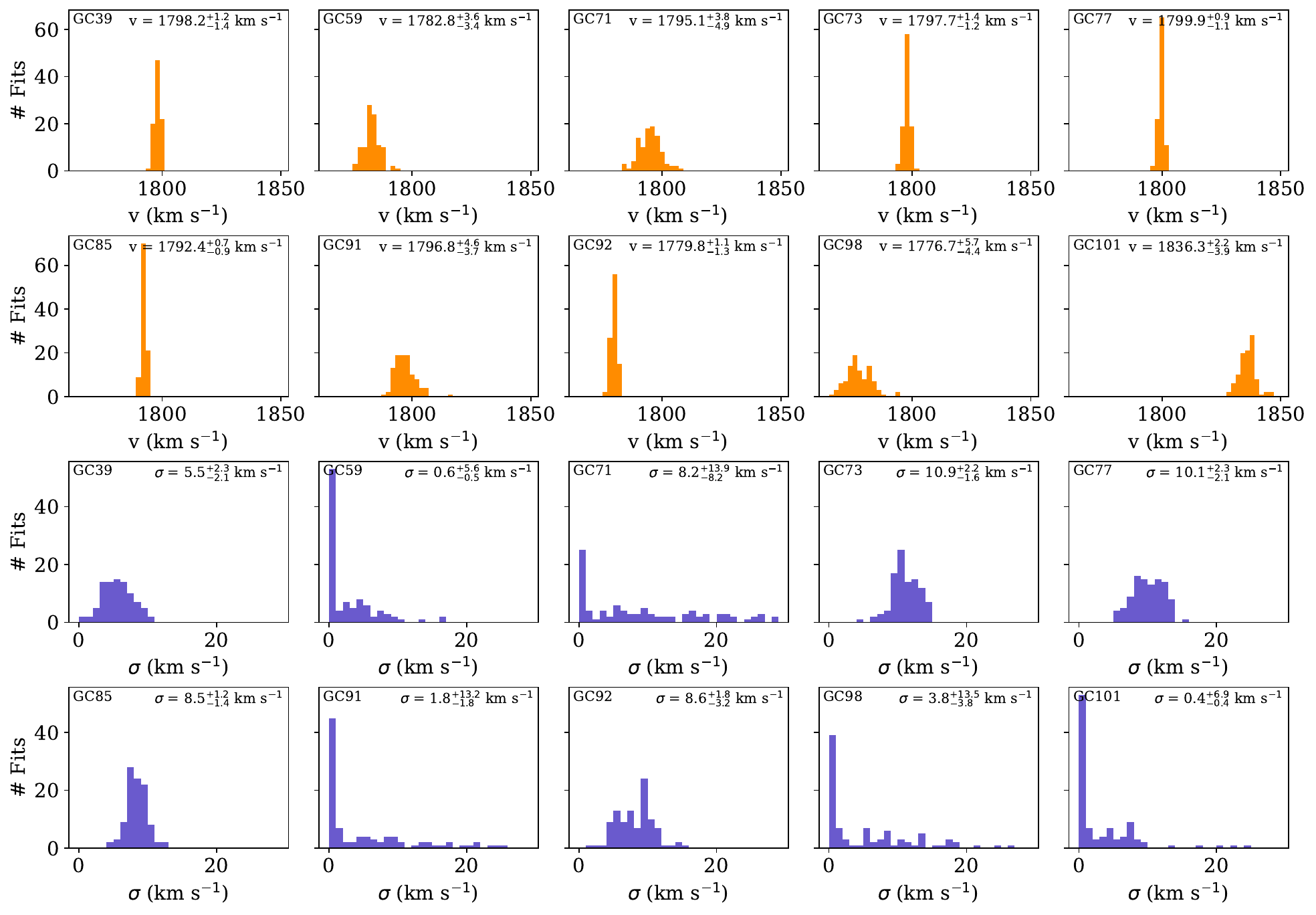}
      \caption{Distributions of radial velocities and velocity dispersions from Monte Carlo of the \textsc{pPXF} results. Velocity dispersions whose distributions show a peak at 0 km s$^{-1}$ indicate that no reliable dispersion is found and are disregarded.
              }
         \label{appendix_Fig2}
   \end{figure}

Figs.~\ref{appendix_Fig1} and \ref{appendix_Fig2} show our fits to individual FLAMES spectra and the resulting velocity and velocity dispersion solutions, respectively.

\FloatBarrier

\section{Aperture and fibre centring corrections}
\label{appendix_B}

\begin{figure}
   \includegraphics[width=\hsize]{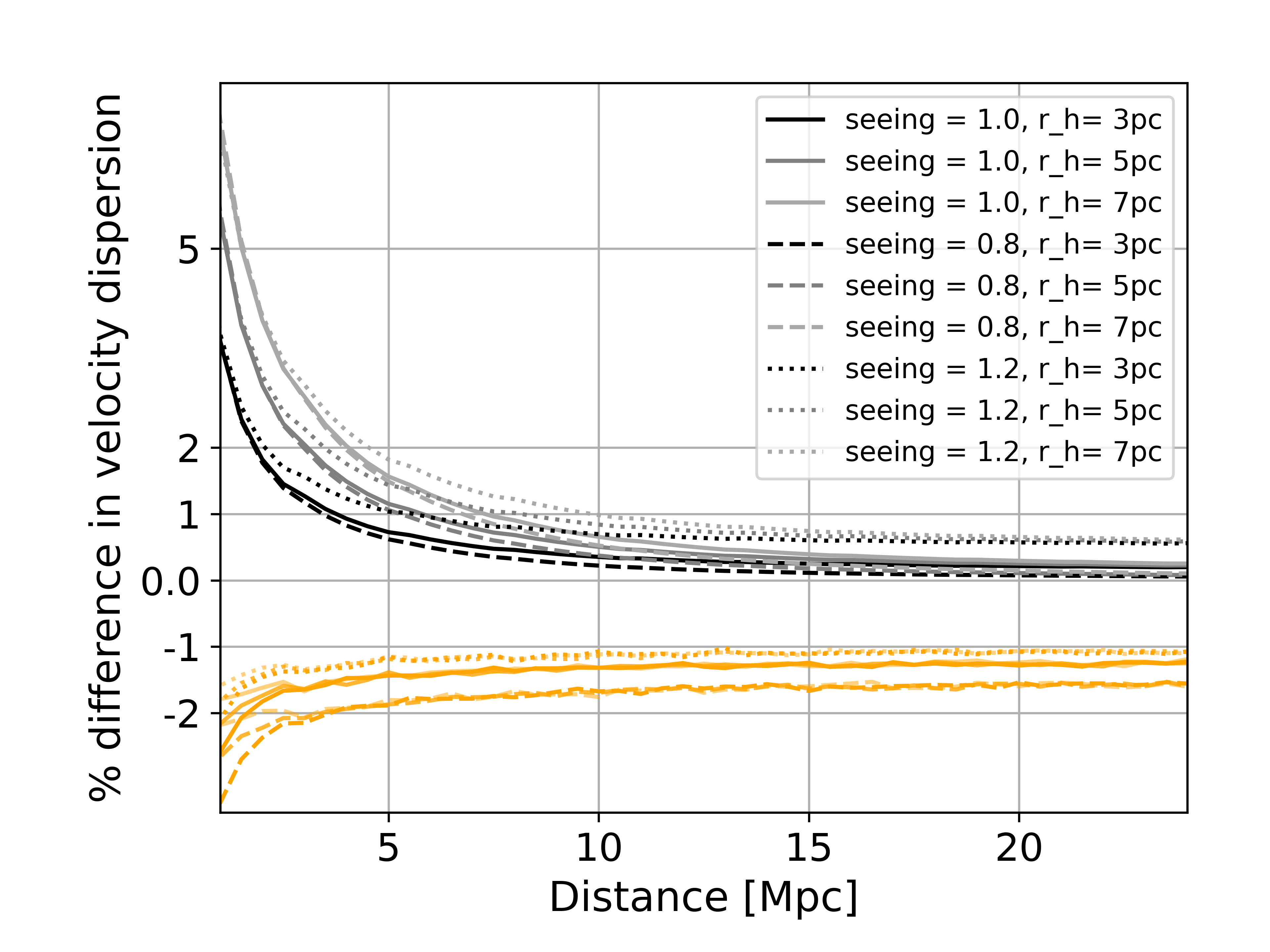}
      \caption{The effect on the measured intrinsic global GC velocity dispersions due to the finite FLAMES fibre size (1.2\arcsec diameter) and fibre positioning error ($\pm0.1$\arcsec) for a range of values for distance, seeing and intrinsic GC size. Positive values indicate an overestimation of the true dispersion due to finite aperture (black lines), while negative values indicate an underestimation of the velocity dispersion due to fibre centring (orange lines).} 
         \label{appendix_Fig3}
   \end{figure}

To estimate aperture effects -- i.e., the impact of missing light from the outer regions of the GCs -- we modelled globular clusters with modified King profiles \citep{Elson1999} for half light radii ($r_h= 3,5,7$ pc). We constructed velocity dispersion profiles using the Jeans equation ($\sigma^2 = G * M(r) / r$)  assuming spherical symmetry and that mass follows light. We convolved the observed profiles with a gaussian with the appropriate FWHM for the range of seeing comparable to our observations (0.8, 1.0 and 1.2$\arcsec$) and then compared the resulting dispersion profiles measured for “infinite” aperture and an aperture of 1.2$\arcsec$, which is the diameter of the FLAMES fibres. These are expressed as the percentage difference from the true values and appear as positive values in Fig.~\ref{appendix_Fig3}, indicating an overestimation in the true velocity dispersion. The aperture correction has a  dependence on distance until about $\sim10$ Mpc. Beyond this, dispersions are overestimated by $<1\%$ even for very large GCs in the poorest seeing. The correction  becomes more significant for closer systems and for larger GCs.

To model the effect of fibre positioning error we ran Monte Carlo simulations which shift the centre of the aperture with respect to the same model GCs assuming an uncertainty of $\pm0.1$\arcsec. This value was chosen based on the FLAMES manual (ESO-281173) which suggests a fibre positioning error of order 0.08\arcsec, combined in quadrature with the GC astrometric error (of order 0.03\arcsec) from DECALs (tied to Gaia DR2).  These are shown as negative values in Fig.~\ref{appendix_Fig3}, indicating an underestimation of the true velocity dispersion. The effect of the uncertainty in the fibre centring works in the opposite sense to the aperture effect and becomes relatively stable with distance beyond $\sim5$ Mpc at the $1-2\%$ level. Again, this effect becomes more important for nearer systems, but more strongly affects smaller clusters (when nearby), which in this case can be understood since their light is more concentrated within the aperture and so any centring uncertainties have a greater effect. 

For our distance of 16.2 Mpc, with seeing = 1.0\arcsec\ (the approximate mean of seeing of our OBs) the aperture variation is $0.25\%$, while the fibre centring variation is $-1.28\%$. This implies that the measured velocity dispersions are likely overestimated at the $\sim1\%$ level. This remains true if we assume the further distance of 21.7 Mpc.

\end{appendix}

\end{document}